\documentclass[journal,twoside,web]{ieeecolor}
\usepackage{tmi}
\usepackage{cite}
\usepackage{amsmath,amssymb,amsfonts}
\usepackage{algorithmic}
\usepackage{graphicx}
\usepackage{textcomp}
\usepackage{color}
\usepackage{multirow, makecell}

\def\BibTeX{{\rm B\kern-.05em{\sc i\kern-.025em b}\kern-.08em
    T\kern-.1667em\lower.7ex\hbox{E}\kern-.125emX}}
\begin{document}

\title{RTNet: Relation Transformer Network for Diabetic Retinopathy Multi-lesion Segmentation}
\author{Shiqi Huang, Jianan Li, Yuze Xiao, Ning Shen and Tingfa Xu
\thanks{This work was supported by the Key Laboratory Foundation under Grant TCGZ2020C004 and Grant 202020429036.}
\thanks{Shiqi Huang, Jianan Li, Yuze Xiao, Ning Shen and Tingfa Xu are with Beijing Institute of Technology, China. Tingfa Xu is also with Chongqing Innovation Center, Beijing Institute of Technology, China (e-mail: huangsq, ciom\_xtf1, lijianan@bit.edu.cn). }
\thanks{Corresponding authors: Tingfa Xu and Jianan Li. }}

\maketitle

\begin{abstract}
Automatic diabetic retinopathy (DR) lesions segmentation makes great sense of assisting ophthalmologists in diagnosis. 
Although many researches have been conducted on this task, most prior works paid too much attention to the designs of networks instead of considering the pathological association for lesions. Through investigating the pathogenic causes of DR lesions in advance, we found that certain lesions are closed to specific vessels and present relative patterns to each other. Motivated by the observation, we propose a relation transformer block (RTB) to incorporate attention mechanisms at two main levels: a self-attention transformer exploits global dependencies among lesion features, while a cross-attention transformer allows interactions between lesion and vessel features by integrating valuable vascular information to alleviate ambiguity in lesion detection caused by complex fundus structures.
In addition, to capture the small lesion patterns first, we propose a global transformer block (GTB) which preserves detailed information in deep network. 
By integrating the above blocks of dual-branches, our network segments the four kinds of lesions simultaneously. 
Comprehensive experiments on IDRiD and DDR datasets well demonstrate the superiority of our approach, which achieves competitive performance compared to state-of-the-arts.

\end{abstract}


\section{Introduction}
\label{sec:introduction}

\IEEEPARstart{D}{iabetic} 
retinopathy (DR) has become a worldwide major medical concern for the large population of diabetic patients and has been the leading cause of blindness in the working-age population today\cite{thomas2019idf,ciulla2003diabetic,raman2016diabetic}. DR lesions often present as microaneurysms (MAs), hemorrhages (HEs), soft exudates (SEs), and hard exudates (EXs) which can be observed in colorful fundus images and are the basis of diagnosis for ophthalmologists. 

However, until now there has been no valid treatment to cure this disease completely. The most recognized treatment is the early diagnosis and intervention to controll the progression of the disease and to avoid eventual loss of vision\cite{wong2018guidelines}. Thus, many national health institutions are promoting DR screening, which has been proven effective in reducing the rate of blindness caused by DR\cite{ciulla2003diabetic,ting2016diabetic}. However, screening is a heavy burden for primary care systems during the promotion, since the ophthalmologists are in very short supply and have already engaged in post-DR treatment. For this reason, automatic segmentation technology for DR lesions has become a trend towards assisting ophthalmologists in diagnosis.

\begin{figure}[!t]
\centerline{\includegraphics[width=\columnwidth]{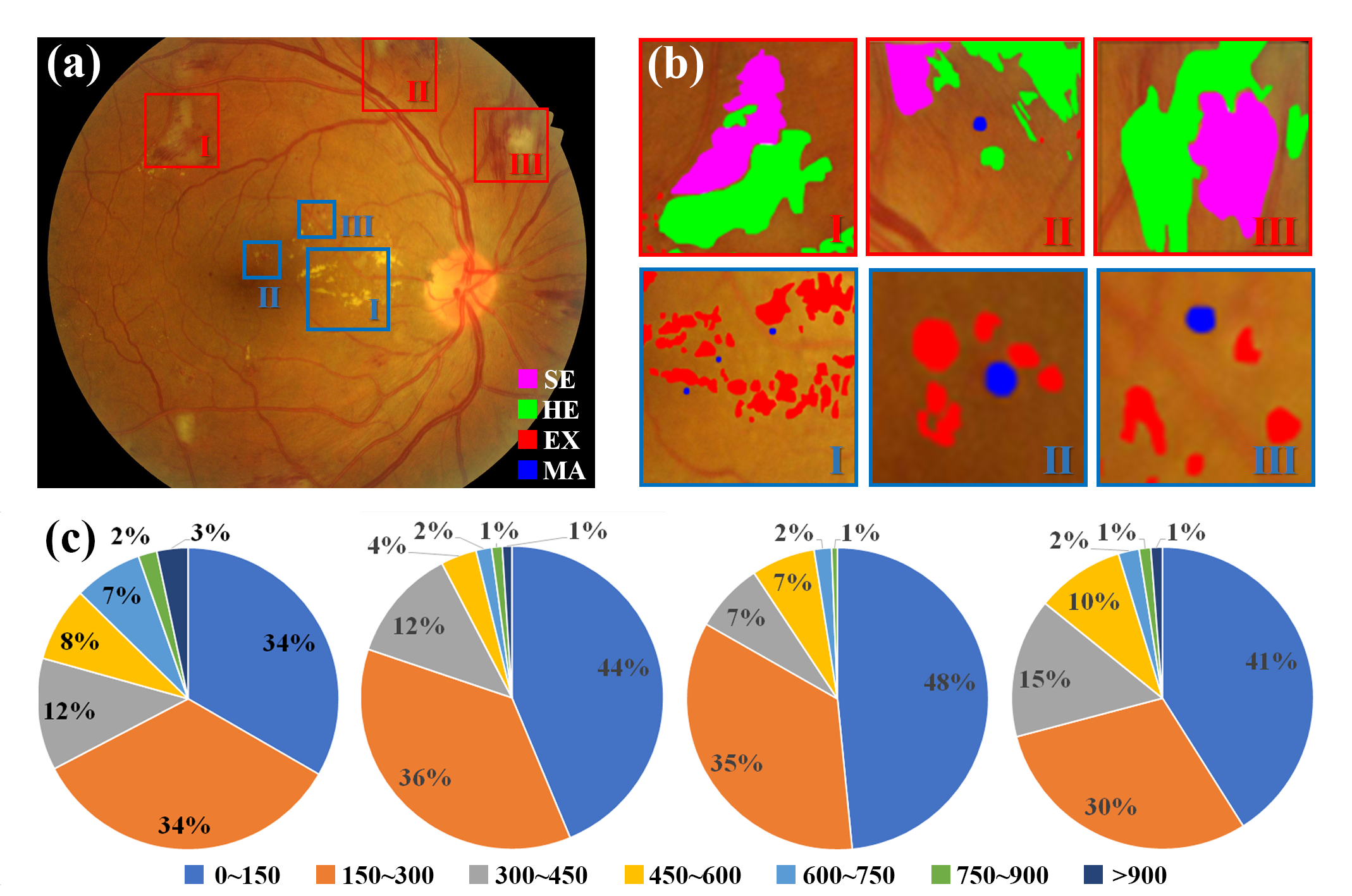}}
\caption{Illustration of fundus image with characteristics of DR lesion. 
SE: soft exudate; HE: hemorrhage; EX: hard exudate; MA: microaneurysm.
(a) Original image with different clusters of lesions denoted by red and blue bounding boxes;
(b) magnified lesion regions where green, purple, red and blue area represent SE, HE, EX and MA, respectively;
(c) location statistics of certain lesions on IDRiD dataset in pixels. Specifically, from left to right are the distances from SE center to nearest HE center, from EX cluster center to nearest MA center, from SE center to the nearest vascular tree midline, and from MA center to the nearest vascular tree midline.}
\label{fig1}
\end{figure}

Recent research efforts have been directed towards automatic DR segmentation based on deep learning (DL). 
Guo et al. \cite{guo2019seg} adopted a pretrained Vgg16\cite{simonyan2014very} as backbone with multi-scale feature fusion block for DR lesion segmentation. Specifically, they extracted side features from each convolution layer in Vgg16 and fused them in a weighted way. Further, a multi-channel bin loss was proposed to alleviate class-imbalance and loss-imbalance problems.
Zhou et al. \cite{zhou2019collaborative} designed a collaborative learning network to jointly improve the performance of DR grading and DR lesion segmentation with attention mechanism. The attention mechanism allowed features with image-level annotations to be refined by class-specific information, and generated pixel-level pseudo-masks for the training of segmentation model.
However, previously published studies paid too much attention to the designs of networks and many of the researches up to now have only achieved the segmentation of just one or two lesions \cite{tavakoli2020automated,mamilla2017extraction,wu2017automatic,khojasteh2019novel,mo2018exudate}. It should be noted that as a complex medical lesion segmentation task, the pathological connections have not received enough attention.

After comprehensive investigation in possible causes of DR lesions, we found two interesting presentation phenomena shown in Fig.\ref{fig1}: 1) lesions are usually closed to specific veins and arteries. For example, SEs are generally distributed at the margins of the main trunk of the upper and lower arteries, and MAs are generally distributed at the margins of the capillaries; 2) most lesions have certain spatial interactions with each other. Specifically, SEs commonly appear at the edge of HEs, while EXs are usually arranged in a circular pattern around one or several MAs, which is consistent with the occurrence of pathology.

Motivated by the above observations, we propose a relation transformer block (RTB), comprised of a cross-attention and a self-attention head, to explore dependencies among lesions and other fundus tissues. For specific, 
the cross-attention head is designed to capture implicit relations between lesions and vessels. We design a dual-branch network to employ both lesion and vascular information, and cross-attention head is integrated between the two branches to make effective use of vascular information in lesion segmentation branch. 
As we know, the fundus tissues, \textit{e.g.,} vessels, optic disc, nerves and other lesions, are complex and easily confused with DR lesions of interest, but considering that vascular information describes certain distribution patterns of the above tissues, the cross-attention head is able to locate more lesions through the layout provided and eliminate the false positives far from certain vessels.
The self-attention head is employed to investigate the relationships of multi-lesion themselves. Since some of the lesions look similar, \textit{e.g.,} HE and MA (both red-lesion), SE and EX (both exduate-lesion), misclassifications between lesions occur frequently. Through impactful information emphasized by self-attention head, the distinction and connection of lesions play roles in reducing confusion in segmentation.
Note that before DL was utilized in medical imaging, vessels were easily mistaken for red lesions, and vessel detection was regarded as a routine step in the segmentation with detected vessels being removed straight away. However, as no fundus dataset is annotated with both vessels and DR lesions, the DL-based lesion segmentation task no longer extracts vessel features separately. To our best knowledge, this is the first trial that utilizes vascular information for deep based fundus lesion segmentation.

In addition, some special lesion patterns, such as MA with small size and SE with blurred border, are hard to be situated accurately due to the lack of fine-grained details in high-level features.
To alleviate this, we propose a Global Transformer Block (GTB) inspired by GCNet\cite{GCNet} to further extract detail information, which can preserve the detailed lesion information and suppress the less useful information of channels in each position. In our network, GTB is also adopted to generate dual branches. Specifically, after the backbone, the shared fundus features are obtain as input of the two GTBs, and then GTBs generate more specific features of vessels and lesions respectively, which would be further investigated at the pathological connection level by RTB.

We have evaluated our network on two publicly available datasets - IDRiD and DDR. Experimental results show that our network outperforms the state-of-the-art DR lesion segmentation reports and achieves the best performance in EX, MA and SE. Furthermore, we also implement ablative experiments on IDRiD dataset and validate the effectiveness of RTB and GTB in improving DR lesion segmentation outcomes.

In summary, our contributions are as follows:
\begin{itemize}
    \item We propose a dual-branch architecture to obtain vascular information, which contributes to locate the position of DR lesions. For effective use of vascular information in multi-lesion segmentation, we design a relation transformer block (RTB) based on transformer mechanism. To our best knowledge, this is the first work to employ multi-heads transformer structure in lesion segmentation in fundus medical images.
    \item We present global transformer block (GTB) and relation transformer block (RTB) to detect the special medical patterns with small size or blurred border. The design explores the internal relationship between DR lesions which improves the performance in capture the details of interest.
    \item Experiments on the IDRiD dataset show that our method achieves a front row finish on DR multi-lesion segmentation. Specifically, our method achieves the best performance in exduates segmentation and ranks second in HE lesion segmentation.
    Experiments on the DDR dataset show that our method outperforms other methods on EX, MA and SE segmentation task and ranks second on HE segmentation task.
\end{itemize}

\section{Related Work} 
\subsection{Pathological Analysis of the DR Lesions}
DR lesions segmentation is a complex topic due to large intra-class variance.
Furthermore, DR lesions vary with different stages of disease as well, which also brings challenges to segmentation. 
However, instead of discovering lesions directly, we notice that there are pathological associations between these lesions, which can be depicted in the spatial distribution. 

We first investigate the possible pathological causes of DR lesions. Briefly, MA is the earliest lesion of DR observed as spherical lateralized swelling which is produced by vascular atresia; EX looks like yellowish-white well-defined waxy patch, generally thought to be lipid produced by the rupture of the retinal nerve tissues, incidentally, which is also resulted from the vascular atresia. Additionally, when the rupture of vessels happens after the vascular atresia, bloods leak out from the vessels, which leads to the lipoproteins in vessels leaking into the retina as well. The leaking bloods form the HE patterns and the leaking lipoproteins form the SE patterns  with poorly defined borders. In summary, as shown in Fig.\ref{fig1}(b), most of the EXs are observed arranged in a circular pattern around one or several MAs (the bottom line) and most of the SEs appear at the edge of HEs (the upper line).

In addition to the intra-class dependencies among lesions, the inter-class relations between DR lesions and vessels also make great sense. As mentioned above, vascular abnormalities are the direct or indirect causes of DR lesions, specifically, we found the fact that SEs are often distributed near the trunk of upper and lower arteries, and MAs are generally distributed among the capillaries. Furthermore, intricate fundus tissues often confuse the identification of lesions, but we notice that there are certain pattern rules of fundus, especially in the distribution of various veins and arteries, which would provide valuable prior information.  

To confirm the above pathological analysis, we count the distances in pixel between different fundus tissues on the IDRiD dataset. Fig.\ref{fig1} (c) illustrates the distance between cluster center of EX and the nearest neighbor MA, SE and the nearest neighbor HE, MA and the closet capillaries, SE and closest upper and lower arteries, respectively. The statistical results verify that there is an exploitable pattern in the distribution of DR lesions. 

\subsection{Deep Neural Networks in DR Lesion Segmentations}

DR lesion segmentation task based on traditional image processing techniques\cite{walter2007automatic,Automatic2005,alipour2012analysis} is facing two main challenges: the great morphological differences of the same lesions in different disease stages and the confusions of DR lesions and similar structures in the fundus. These two problems have not been effectively solved until 
deep neural networks (DNNs) exploded in the field of computer vision (CV) \cite{krizhevsky2012imagenet} and have also been widely applied to DR lesion segmentations\cite{CABNet,CANet,CENet}.

However, DNNs also raise new difficulties. For instance, the detailed information is easily lost by deep networks but most of the DR lesions are very small and even just one or two pixels. Besides, considering the balance of different characteristics of red and exudate lesions, the accuracy of multi-task model is limited. 

To deal with the above issues, researchers have proposed many improvements, which can be summarised as two directions:

Firstly, some researchers focus on the designs of attention models fusing low-level and high-level features together to avoid details lost in the deep network.
Zhang et al. \cite{zhang2019detection} fused multiple features with distinct target features in each layer based on attention mechanism and achieved preliminary MA detection.
Wang et al. \cite{wang2017zoom} designed a dual-branch attention network, with one producing a 5-graded score map and the other producing an attention gate map combined to the score map to highlight suspicious regions.
Zhou et al. \cite{zhou2019collaborative}  applied low-level and high-level guidances to different lesion features and obtained the  refined multi-lesion attention maps, which were further employed as pseudo-masks to train the segmentation model.

Secondly, the task of segmenting DR lesions is divided into segmenting red lesions and exudate lesions separately, which evades the balanced cost of inter-class disparities and enables fully learning of the same type of lesions.
Mo et al. \cite{mo2018exudate} designed a fully convolutional residual network incorporating multi-level hierarchical information to segment the exudates without taking the red lesions into account.
Xie et al. \cite{xie2020sesv} built a general framework to predict the errors generated by existing models and then correct them, which performed well in MA segmentation.

However, the first direction pays much attention to the network designs, with seldom considering the pathological connections of DR lesions, while the second direction leads to time consuming and large memory requirements. In order to take advantages of the pathological connections and improve the efficiency of the multi-task model, we propose a RTB consisting of self-attention and cross-attention head to segment DR lesions simultaneously.

\subsection{Transformer in Medical Images}

Transformer network has been one of the fundamental architecture for natural language processing (NLP) since 2017 due to the efficient and effective self-attention mechanism\cite{vaswani2017attention}. It improves the performance on many NLP tasks, such as text classification, general language understanding and question answering. Compared with recurrent networks, transformer network achieves parallel computation and reduces the computational complexity. 
In a basic transformer attention block, Query ($Q$), Key ($K$), Value ($V$) are the three typical inputs to a attention operation. At first, $Q$ and $K$ are  computed in the form of pairwise function to obtain the corresponding attention of all points on $K$ for each point on $Q$. The pairwise function can optionally be Gaussian, Embedding Guassian, Dot-Product, Concatenation and \textit{etc.}. Then, the product is multiplied by $V$ and passes through a column-wise softmax operator to ensure every column sum to 1. Every position on the output contains the recoded global information by attention mechanism. In self-attention operator, $Q = K = V$, so that the output has the same shape with input.

Inspired by the success in the domain of NLP, a standard transformer was applying to CV in 2020\cite{dosovitskiy2020image} with the fewest modifications, called as Vision Transformer (ViT). The input to a ViT is a sequence of cropping images which are linearly encoded by aliquoting the original image. The image patches are treated the same way as tokens (words) in an NLP application.

Recently, the transformer architecture has also been applied to the field of medical image processing. 
 Liu et al. \cite{GPT} proposed a global pixel transformer (GPT) to predict several target fluorescent labels in microscopy images. The GPT is similar to a three-headed transformer with different sizes of query inputs, which allows it to adequately capture features at different scales.
Guo et al. \cite{guo2021transformer} applied the ViT to anisotropic 3D medical image segmentation, with the self-attention model arranged at the bottom of the Unet architecture.
Song et al.\cite{DRT} built a Deep Relation Transformer (DRT) to combine OCT and VF information for glaucoma diagnosis. They modified the standard transformer to an interactive transformer that utilizes a relationship map of VF features interacting with OCT features. 

\section{Methodology}
In this section, we first give a brief overview of our proposed network, and then elaborate on the key network components, \textit{i.e.,} global transformer block (GTB) and relation transformer block (RTB). Finally, designed loss function is further provided.

\subsection{Overview}
\begin{figure*}[htbp]
\centerline{\includegraphics[width=\textwidth]{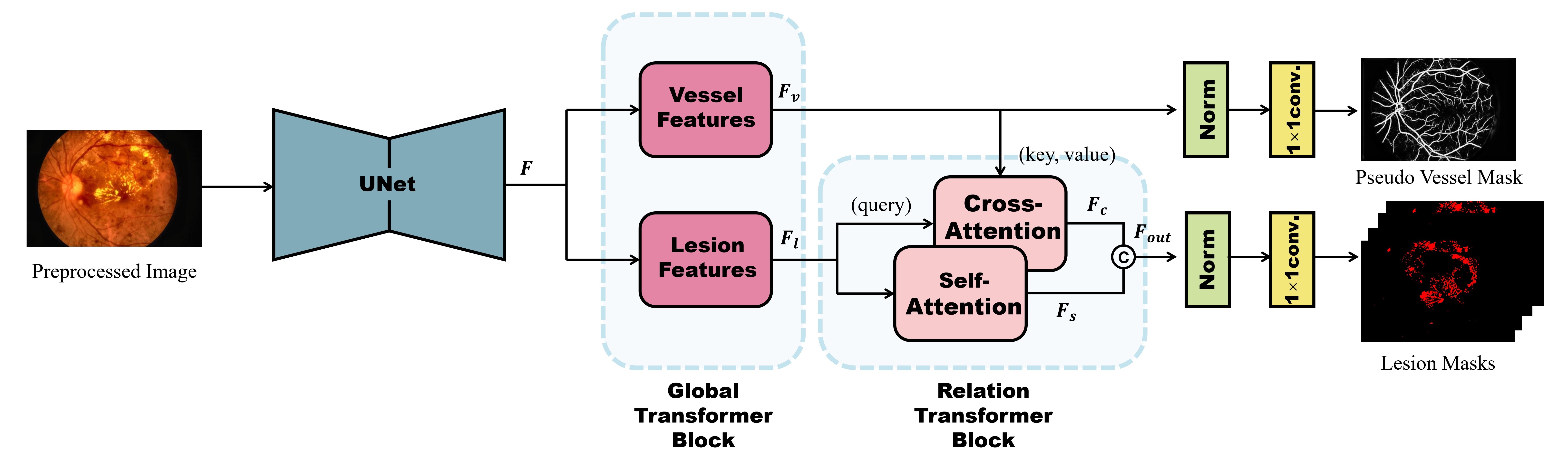}}
\caption{Pipeline of the proposed method. The input image passes through a backbone to obtain the shared feature $\mathbf{F}$. Then the shared feature $\mathbf{F}$ takes two branches to achieve vessels and multi-lesion segmentation respectively. Two Global Transformer Blocks (GTB) are applied to both branches to generate specific features, and a Relation Transformer Block (RTB) is incorporated after GTB to explore the inherent pathological connections among multi-lesion and between multi-lesion and vessels.}
\label{fig2}
\end{figure*}

Given an input fundus image, the proposed network is designed to output one vascular mask and four lesion masks in parallel.
Fig.\ref{fig2} depicts its overall architecture, which is comprised of four key components: backbone, global transformer block (GTB), relation transformer block (RTB), and segmentation head.
A dual-branch architecture is employed upon the backbone to explore vascular and pathological features separately, where the transformers based on GTB and RTB
are incorporated to reason about interactions among both features.

To be specific, the fundus image first passes through a backbone to obtain an abstracted feature map $\mathbf{F}$, with a spatial resolution of $W\times H$ and $C$ number of channels.
Then, two parallel branches comprised of global transformer block (GTB) are incorporated to exploit long-range dependencies among pixels in $\mathbf{F}$, resulting in specific vessel features $\mathbf{F}_v$ and primary lesion features $\mathbf{F}_l$ fueled with global contextual information, respectively.
Upon the branch providing lesion features, we further integrate a relation transformer block (RTB) to model spatial relations between vessels and lesions due to their inherent pathological connections using a self-attention and a cross-attention head:
the self-attention head inputs only the lesion features $\mathbf{F}_l$, and exploits long-range contextual information to generate self-attentive features $\mathbf{F}_{s}$ through a self-attention mechanism; the cross-attention head inputs both the lesion and vessel features $\mathbf{F}_{l}$, $\mathbf{F}_v$,
and incorporates beneficial fine-grained vessel structural information into $\mathbf{F}_v$, producing cross-attentive features $\mathbf{F}_{c}$. 
The resulting $\mathbf{F}_{s}$ and $\mathbf{F}_{c}$ are concatenated together to form the output of the RTB. 
Finally, two sibling heads, each of which contains a Norm layer and a $1\times 1$ convolution, are used to predict vascular and pathology masks based on the vessel features and concatenated lesion features, respectively.

GTB contains one head while RTB contains two heads. Although the basic heads of GTB and RTB are based on the transformer structure that generates query, key and value for relation reasoning,
they are structurally different in our work. 
The query of head in GTB is similar to a channel-wise weights and the one in RTB has the same size in spatial dimension with the input.
In a training process, GTB is employed to generate specific multi-lesion and vessel features independently which maintain more details of interest, and RTB further exploits the inherent pathogenic relationships between multi-lesion and vessels, which eliminate noise and imply the location information. 

\begin{figure}[htbp]
\centerline{\includegraphics[width=\columnwidth]{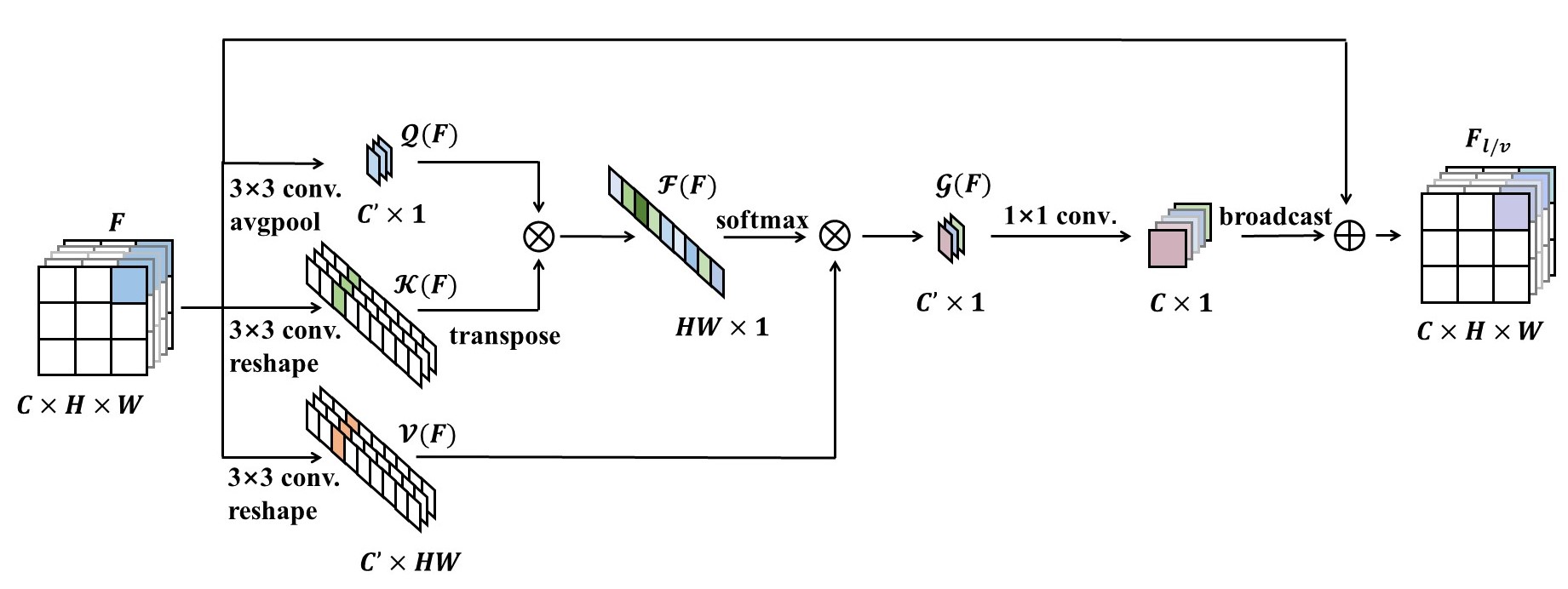}}
\caption{The overall structure of Global Transformer Block (GTB).}
\label{fig3}
\end{figure}

\begin{figure}[htbp]
\centerline{\includegraphics[width=\columnwidth]{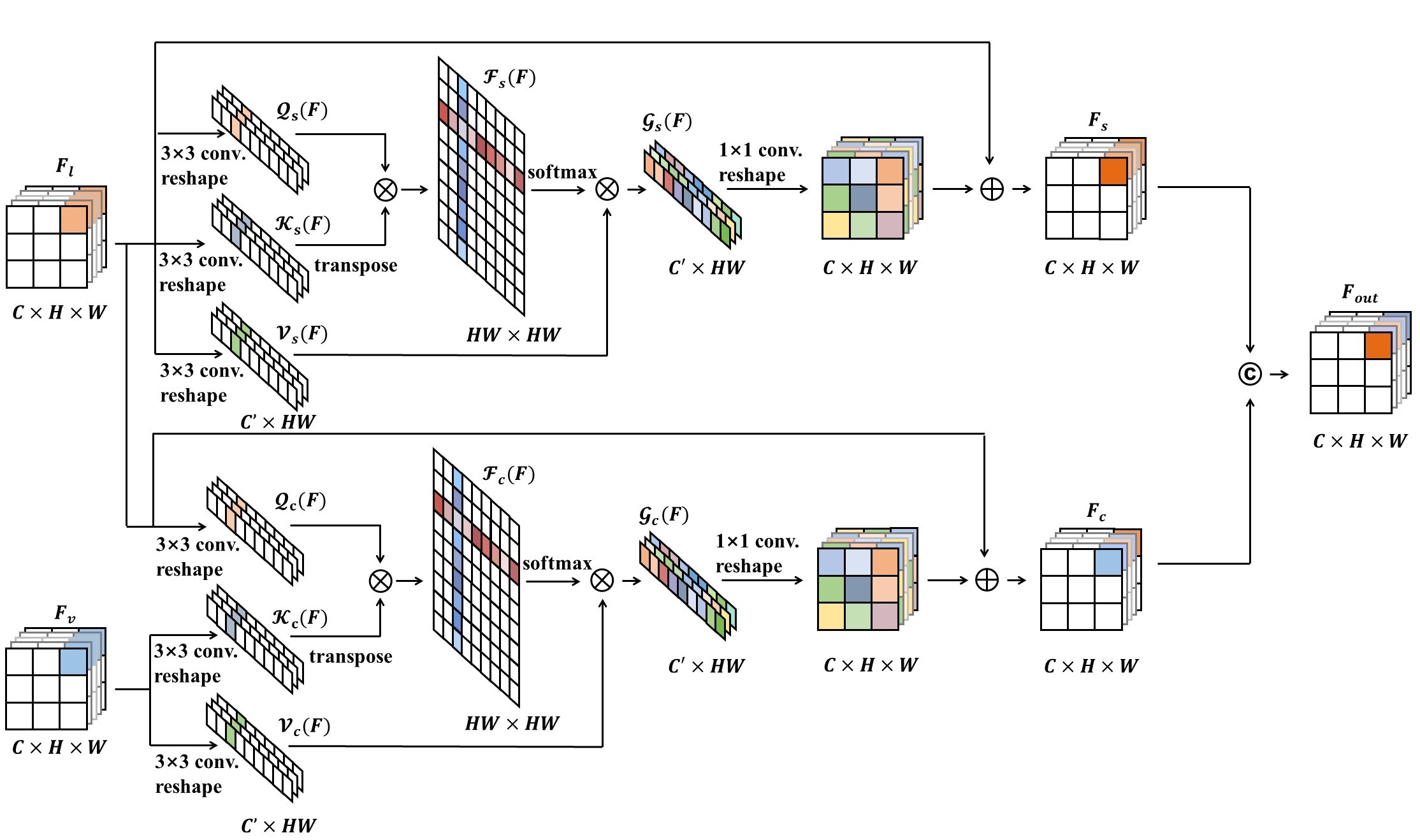}}
\caption{The details of Relation Transformer Block (RTB).}
\label{fig4}
\end{figure}

\subsection{Global Transformer Block}
The Global Transformer Block (GTB) contains two parallel branches of the same architecture to extract features for lesions and vessels separately. Such a dual-branch design owns to the fact that lesions and vessels generally have dramatically different visual patterns. To be concrete, lesions are discrete patterns, with nearly random spatial distribution. While vessels are topological connected structures, and the layout of vascular trunks, containing central retinal artery, ciliary artery and etc, generally follows some common rules. 
It is hence necessary to use specialized branches to learn specific characteristics of different objects of interest.

Fig.\ref{fig3} presents the detailed structure of each GTB branch.
It takes an input $\mathbf{F}\in \mathbb{R}^{\: C\times W\times H}$ generated from the backbone, and outputs attentively-refined feature maps $\mathbf{F}_{i}\in \mathbb{R}^{\: C\times W\times H},i \in\left\{l,v\right\}$ of lesions and vessels respectively.
Specifically, GTB follows the typical framework of transformer networks. 
Three generators, denoted as $\mathcal{Q}$, $\mathcal{K}$ and $\mathcal{V}$ are first employed to transform the input $\mathbf{F}$ into query, key and value, respectively. In GTB, the generator  $\mathcal{Q}$ is implemented with a $3 \times 3$ convolution followed by a global average pooling, and outputs a query vector $\mathcal{Q}(\mathbf{F}) \in \mathbb{R}^{\: C^{'}\times 1}$ with the channel number designed as $C^{\prime}=C/8$; the generator $\mathcal{K}$ and $\mathcal{V}$ have the same architecture as $\mathcal{Q}$ expect for replacing the global averaging pooling with a reshape operation, leading to the key and value $\mathcal{K}(\mathbf{F}),\mathcal{V}(\mathbf{F})\in \mathbb{R}^{\:C^{'}\times HW}$.

We define the pairwise function of query and key as a matrix multiplication:
\begin{equation}
    \mathcal{F}(\mathbf{F})=\mathcal{K}(\mathbf{F})^T\mathcal{Q}(\mathbf{F}) ,
\end{equation}
where the superscript $T$ denotes a transpose operator for matrix. 
Note that the query of GTB acts as a channel-wise query instead of the position query as NLNet\cite{wang2018NLnet}. To be specific, the query vector is considered as a feature selector for channels of key matrix.  
Subsequently, the product $\mathcal{F}(\mathbf{F})\in\mathbb{R}^{\:HW\times 1}$ also acts as a feature selector for spatial positions of value matrix. 
In summary, the GTB can be roughly described as a attention mechanism which fuses channel-wise first and then spatial-wise weighted features together with input information.

Next, we consider the global transform operation defined as:
\begin{equation}
    \mathcal{G}(\mathbf{F}) = \mathcal{V}(\mathbf{F})softmax(\mathcal{F}(\mathbf{F}))\in \mathbb{R}^{\:C^{'}\times 1},
\end{equation}
where $softmax$ is a softmax function to normalize the $\mathcal{F}(\mathbf{F})$.

Then, We take the obtained attentive features $\mathcal{G}(\mathbf{F})$ with a linear embedding as a residual term to the input $\mathbf{F}$, and get the final output through a residual connection:
\begin{equation}
    \mathbf{F}_{i} = W\mathcal{G}(\mathbf{F})+\mathbf{F},\ i\in\left \{ l,v\right \},
\end{equation}
where the $+$ operation denotes broadcasting element-wise sum operation; the $W$ is a linear embedding, implemented as $1 \times 1$ convolution to convert the channel number of intermediate feature map from $C^{'}$ back to $C$. As a result, the output features are received with the same format as the input, but have been enriched with specialized vessel and lesion features, respectively.

The GTB structure is inspired by the GCNet\cite{GCNet}. They both follow the idea of transformer mechanism but generate the per-channel weights. The weight vectors in GCNet and GTB both obtained by a matrix multiplication, but different from GCNet, GTB attains the two multipliers with three generator to further highlight the useful channels in each position, realizing both channel-wise and spatial-wise attention. Due to the fundus lesion features, especially the small discrete ones, are easily confused with artifacts or idiosyncratic tissues, 
useful information tends to exist in only a few pixels of certain channels. It is an improved method aimed at the feasibility of small discrete pattern segmentation in fundus images.

\subsection{Relation Transformer Block}
Relation Transformer Block (RTB) consists of a self-attention and a cross-attention head, used to capture intra-class dependencies among lesions and inter-class relations between lesions and vessels, respectively, as shown in Fig.\ref{fig4}. In each head, three trainable linear embeddings, implemented with a $3 \times 3$ convolution followed by a reshape operation, are employed as the query, key and value generator $\mathcal{G}_i, \mathcal{K}_i, \mathcal{V}_i,i \in\left\{s,c\right\}$, respectively.
The pairwise computations of query and key in self-attention head and cross-attention head are described as:
\begin{equation}
    \begin{split}
        \mathcal{F}_{s}(\mathbf{F}_l)&=\mathcal{K}_{s}(\mathbf{F}_l)^T\mathcal{Q}_{s}(\mathbf{F}_l)  \\
        \mathcal{F}_{c}(\mathbf{F}_l,\mathbf{F}_v)&=\mathcal{K}_{c}(\mathbf{F}_v)^T\mathcal{Q}_{c}(\mathbf{F}_l),
    \end{split}
\end{equation}
where the subscripts $s$ and $c$ denote the self-attention and the cross-attention head, respectively.
It is important to emphasize that different from the self-attention head that derives the query, key all from input lesion features $\mathbf{F}_l$, the cross-attention head generates the key from the vessel features $\mathbf{F}_v$ instead to integrate vascular information.

Next, the individual attentive features of the two heads are computed respectively as:
\begin{equation}
\begin{split}
    \mathcal{G}_{s}(\mathbf{F}_{l})&=\mathcal{V}_{s}(\mathbf{F}_{l})softmax(\mathcal{F}_{s}(\mathbf{F}_{l}))\\
    \mathcal{G}_{c}(\mathbf{F}_{l},\mathbf{F}_v)&=\mathcal{V}_{c}(\mathbf{F}_{v})softmax(\mathcal{F}_{c}(\mathbf{F}_{l},\mathbf{F}_v)).
\end{split}
\end{equation}

We adopt residual learning to each head as well and get the outputs:

\begin{equation}
\begin{split}
    \mathbf{F}_{i} = W_{i}\mathcal{G}_{i}(\mathbf{F}_{l},\mathbf{F}_v)\oplus\mathbf{F}_l\\i\in \left\{s, c \right\},
\end{split}
\end{equation}
where the $W_i$ is a linear embedding implemented as $1 \times 1$ convolution, and the $\oplus$ operation is performed by a residual connection of element-wise addition. 

As such, the self-attention head computes the response in a position as a weighted sum of the features in all positions, and thus well captures long-range dependencies. 
Given the fact that DR lesions are usually dispersed over a broad range, the self-attention can exchange message among multiple lesions, regardless of their positional distance, and thus allows the modeling of intra-class pairwise relations of lesions. The head is supposed to distinguish the mixtures of more than two lesions and further refine the edges of large patterns in lesion segmentation.

The cross-attention head queries global vascular structures from the vessel features, thus incorporating interactions between lesions and vessels.
Considering that lesions and vessels have strong inherent pathogenic connections, the cross attention help to better locate MA and SE, and meanwhile eliminate false positives of EX caused by vessel reflection and MA caused by capillary confusion.

We concatenate the resulting features $\mathbf{F}_{s}$ from the self-attention head and that $\mathbf{F}_{c}$ from the cross-attention head, leading to the final RTB output:

\begin{equation}
    \mathbf{F}_{out} = [\mathbf{F}_s;\mathbf{F}_c],
\end{equation}
where the $[\cdot\ ;\ \cdot]$ denotes the concatenation at channel dimension.

\subsection{Loss Function}
We employ two loss functions, \textit{i.e.}, $\mathcal{L}_{lesion}$ and $\mathcal{L}_{vessel}$ for the multi-lesion and vessel segmentation branches respectively, and the total loss of our network is defined as:
\begin{equation}
    \mathcal{L} =\mathcal{L}_{lesion}+\lambda \mathcal{L}_{vessel},
\end{equation}
where the $\mathcal{L}_{lesion}$ denotes the 5-class weighted cross-entropy loss for multi-lesion segmentation and the $\mathcal{L}_{vessel}$ is a binary weighted cross-entropy loss to learn vascular features; the $\lambda$ is set as the weight in the loss function. When $\lambda = 0.0$, the network is optimized by the multi-lesion features only, and as $\lambda$ grows, vascular information plays an increasing role in optimization.

\section{Experiments And Results}
\subsection{Datasets}
\textbf{IDRiD Dataset} is available for the segmentation and grading of retinal image challenge 2018\cite{porwal2018indian, porwal2020idrid}. The segmentation part of the dataset contains 81 $4288 \times 2848$ sized fundus images, accompanied by four pixel-level annotations, \textit{i.e.,} EX, HE, MA and SE if the image has this type of lesion. In total, there are 81 EX annotations, 81 MA annotations, 80 HE annotations, and 40 SE annotations. The partition of training set and testing set is provided on IDRiD already, with 54 images for training and the rest 27 images for testing. 

\textbf{DDR Dataset} is provided by Ocular Disease Intelligent Recognition (ODIR-2019) for lesion segmentation and lesion detection\cite{LI2019}. This dataset consists 13,673 fundus images from 147 hospitals, covering 23 provinces in China. For segmentation task, 757 fundus images are provided with pixel-level annotation for EX, HE, MA and SE if the image has this type of lesion. In total, there are 486 EX annotations, 570 MA annotations, 601 HE annotations, and 239 SE annotations. The partition of training set, validation set and testing set is provided on DDR already, with 383 images for training, 149 images for validation and the rest 225 images for testing.

\begin{table*}
    \centering
    \caption{Performance Comparison with the state-of-the-art works reported on the IDRiD dataset, where the \textbf{separate} and \textbf{same} indicates the way in which the method segments the lesions separately by different models or at the same time by one model}
    \label{tab1}
    \begin{tabular}{|c|c|c|c|c|c|c|c|c|c|c|}
    \hline
        \multirowcell{2}{Method} & \multirowcell{2}{sepa-\\rate} & \multirowcell{2}{same}  & \multicolumn{2}{c|}{Hard   Exudates} &\multicolumn{2}{c|}{Haemorrhages} 
        & \multicolumn{2}{c|}{Microaneurysms} 
        & \multicolumn{2}{c|}{Soft   Exudates}  \\ \cline{4-11}
         & & & AUC\_PR & AUC\_ROC & AUC\_PR & AUC\_ROC & AUC\_PR & AUC\_ROC & AUC\_PR & AUC\_ROC \\ \hline
        VRT(1st) & \checkmark & & 0.7127  & - & 0.6804  & - & 0.4951  & - & 0.6995  & - \\ 
        PATech(2nd) & \checkmark & & 0.8850  & - & 0.6490  & - & 0.4740  & - & - & - \\ 
        iFLYTEK-MIG(3rd) & \checkmark & & 0.8741  & - & 0.5588  & - & 0.5017  & - & 0.6588  & - \\ 
         
        DRUNet\cite{kou2019microaneurysms} & \checkmark & & - & - & - & -  & - & 0.9820  & - & - \\ 
        SESV\cite{xie2020sesv} & \checkmark & & - & - & - & - & \textbf{0.5099}  & - & - & - \\
        L-Seg\cite{guo2019seg} & & \checkmark & 0.7945  & - & 0.6374  & - & 0.4627  & - & 0.7113  & - \\
        SSCL\cite{zhou2019collaborative} & & \checkmark & 0.8872  & 0.9935  & \textbf{0.6936}  & \textbf{0.9779}  & 0.4960  & 0.9828  & 0.7407  & 0.9936  \\ 
        RTN(Ours) & & \checkmark & \textbf{0.9024}  & \textbf{0.9980}  & 0.6880  & 0.9731  & 0.4897  & \textbf{0.9952}  & \textbf{0.7502}  & \textbf{0.9938}  \\ \hline
    \end{tabular}
\end{table*}

\begin{table*}
\caption{Performance Comparison with state-of-the-art segmentation methods on the DDR dataset, where * denotes the results are reproduced by ourselves}
\label{tab2}
    \centering
    \begin{tabular}{|c|c|c|c|c|c|c|c|c|}
    \hline
        \multirowcell{2}{Method} & \multicolumn{2}{c|}{Hard   Exudates} &\multicolumn{2}{c|}{Haemorrhages} 
        & \multicolumn{2}{c|}{Microaneurysms} 
        & \multicolumn{2}{c|}{Soft   Exudates} \\ \cline{2-9}
         & AUC\_PR & AUC\_ROC & AUC\_PR & AUC\_ROC & AUC\_PR & AUC\_ROC & AUC\_PR & AUC\_ROC \\ \hline
        HED\cite{xie2015holistically} & 0.4252 & 0.9612 & 0.2014 & 0.8878 & 0.0652 &0.9299 & 0.1301 & 0.8215 \\ 
        DeepLab v3+\cite{chen2018deeplabv3} & 0.5405 & 0.9641 & 0.3789 & 0.9308 & 0.0316 & 0.9245 & 0.2185 & 0.8642 \\
        UNet\cite{guan2019fully,Yakubovskiy2019} & 0.5505 & 0.9741 & \textbf{0.3899} & \textbf{0.9387} & 0.0334 & 0.9366 & 0.2455 & 0.8778 \\
        L-seg*\cite{guo2019seg} & 0.5645 & 0.9726 & 0.3588 & 0.9298 & 0.1174 & 0.9423 & 0.2654 & 0.8795\\ 
        RTN(Ours) & \textbf{0.5671} & \textbf{0.9751} & 0.3656 & 0.9321 & \textbf{0.1176} & \textbf{0.9452} & \textbf{0.2943} & \textbf{0.8845} \\ \hline
    \end{tabular}
\end{table*}

\subsection{Implementation Details}

\subsubsection{Data Preparation}
To prepare more trainable data, some operations are performed on the original images. First, the images are input into the a segmentation model pretrained on DRIVE\cite{staal2004ridgeb91} and STARE\cite{hoover2000locatingb92} dataset with vessel annotations and the pseudo vascular masks are obtained. Next, 
limited by the memory received, the large images are random resized and cropped into small pieces of $512 \times 512$ size,  additionally we apply random horizontal flips, vertical flips, and random rotation as forms of data augmentation to reduce overfitting. Then, in order to enhance image contrast while preserving local details, we process Contrast Limited Adaptive Histogram Equalization (CLAHE) on all input images with ClipLimit=2 and GridSize=8 by setting. The CLAHE is proven to be effective due to the anomalousness distinguished from the background in diabetic fundus, and the quantitative results are presented in Table \ref{tab5}.

\subsubsection{Model Settings}
The typical UNet architecture\cite{Yakubovskiy2019} is  
a popular method for medical images segmentation, constructed by an encoder and a decoder with skip connections in channel-wise concatenation manner. In this paper, we apply DenseNet-161\cite{huang2017densely} pretrained on ImageNet dataset as the backbone of UNet encoder\cite{guan2019fully} to achieve better performance. The channel number $C$ of output of UNet is set to 32.

\subsubsection{Experiment settings}
Our framework is implemented using pytorch backend and performed on NVIDIA GeForce RTX 3090 GPU with 24GB of memory. During the training, the batch-size is set to 16. The initial learning rate is set to 0.001 and is decay in a step-wise manner to 0.1 times of the previous every 120 epochs. All models are trained for 250 epochs with the SGD optimizer with momentum 0.9 and weight decay 0.0005.

The loss function settings are as follows: a) the return loss ratio $\lambda$ is set to 0.1; b) the weights of $\mathcal{L}_{lesion}$ are set as 0.001, 0.1, 0.1, 1.0, 0.1 for background, EX, HE, MA and SE respectively; c) The coefficients of background and vessels in $\mathcal{L}_{vessel}$ are set as 0.01 and 1.0.

\subsection{Evaluation Metrics}
To evaluate the performance of the proposed method, we employ the area-under-the-curve (AUC) of both the precision and recall (PR) curve and receiving operating characteristic (ROC) curve \cite{scikit-learn}, which are also recognized as metrics of fundus image segmentation in previous competitions and researches. The former is more concerned with the accuracy of the true data in prediction, while the latter reflects the performance of the data predicted positively. There is more emphasis on recall metric in medical images, which indicates the performance of true samples being successfully predicted, \textit{i.e.,} the AUC\_PR drawn by recall metric shows more practical value, and the AUC\_ROC also characterizes the effectiveness of the model.

\subsection{Comparisons on Other State-of-the-art Methods}
We compare our method with previous works reported on the IDRiD dataset (Table \ref{tab1}). Our method ranks first in AUC\_ROC of EX, MA and SE and AUC\_PR of EX and SE, ranks second in AUC\_ROC and AUC\_PR of HE. 
Note that the first five methods of Table \ref{tab1} all employed individual models segmenting the four lesions separately: according to the conference reports of the top 3 IDRiD competition teams, four models were developed for segmentation of four lesions respectively; DRUNet\cite{kou2019microaneurysms} and SESV\cite{xie2015holistically} proposed a specific network to segment MA only. Although SESV achieved the best AUC\_PR of MA, such individual designs require modifying a large number of hyper-parameters in training stage and lead to time consuming in inference.
The rest methods of Table \ref{tab1}: L-Seg\cite{zhou2019collaborative}, SSCL\cite{zhou2019collaborative} and our network, all propose one model to segment the four lesions at the same time. SSCL performs better than ours in AUC\_PR of MA segmentation but worse in AUC\_ROC, that might result from the fact that although RTB reduces false positives of MA away from the vessels, it also introduces the false negatives of MA.

We also verify the effectiveness of the proposed method on DDR dataset. Compared with the IDRiD dataset, the DDR dataset is an updated dataset with few results reported on it, so we apply some state-of-the-art segmentation methods on the DDR dataset to make comparisons. As shown in Table \ref{tab2}, in the comparison with other state-of-the-arts, our method achieves the best performance in EX, MA and SE and ranks second in HE. It is obvious that our method is not doing the best job in HE segmentation both in IDRiD and DDR dataset, which may be explained by the fact that the large HE patterns are formed by blood irregularly haloing on the retina, and the specific bleeding points have been blurred. In this case, RTB, an approach that is more concerned with theoretical correlations, does not contribute useful information in the segmentation of large HEs. As mentioned in the dataset section, considering there are many low quality images with uneven illumination, underexposure, overexposure, image blurring, retinal artifacts and other disturbing lesion tissues in DDR dataset, DDR dataset is more challenging than IDRiD dataset, which results in the performance of the former lagging behind that of the latter.

\begin{table}[]
  \centering
  \caption{Performance Comparison of the different backbones with our network}
  \scalebox{0.9}{
    \begin{tabular}{|c|c|c|c|c|c|}
    \hline
    \multirowcell{2}{Framework}  &
    \multirowcell{2}{Encoder}&
    \multicolumn{4}{c|}{AUC\_PR}\\\cline{3-6}
      & & EX & HE & MA & SE \\
    \hline
     \multirowcell{5}{UNet\cite{guan2019fully}} & ResNet-34\cite{he2016deep} & 0.8778    & 0.6764  & 0.4659   & 0.7407   \\
     & ResNet-50\cite{he2016deep} & 0.8858    & 0.6874   & 0.4692   & 0.7475   \\
     & Xception\cite{chollet2017xception} & 0.8924     & 0.6806 & 0.4851 & 0.7424 \\
     & Vgg19-bn\cite{simonyan2014very} & 0.8821    & 0.6832  & 0.4789    & 0.7423    \\
     & DenseNet-161\cite{huang2017densely} & \textbf{0.9024}    & \textbf{0.6880}   & \textbf{0.4897}    & \textbf{0.7502}   \\
    \hline
    \end{tabular}}
  \label{tab4}
\end{table}

\subsection{Ablation Studies on IDRiD Dataset}
\begin{figure*}[htbp]
\centerline{\includegraphics[width=\textwidth]{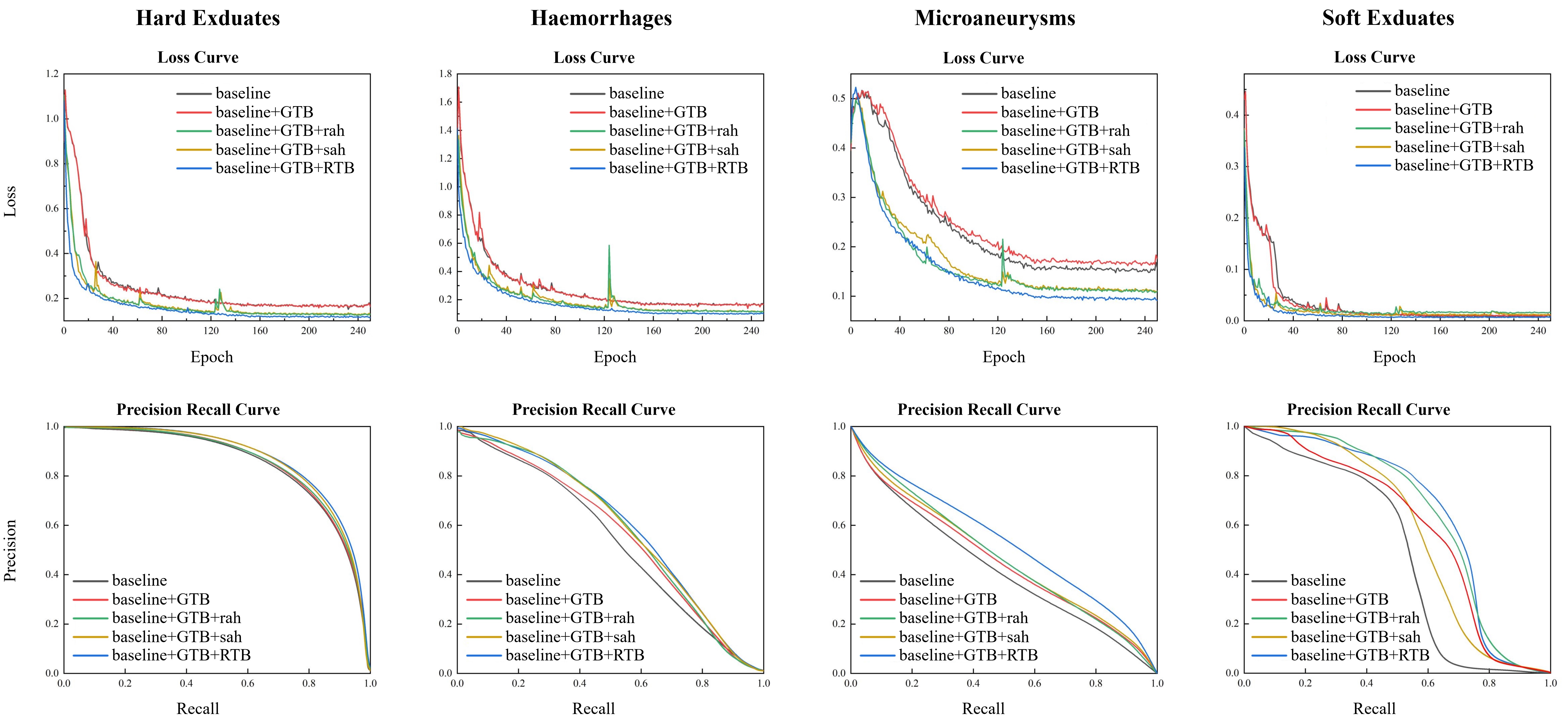}}
\caption{Loss and PR curves for segmentation over four DR lesions. Ablation studies are compared to explore the effectiveness of the baseline itself and stacked by GTB, self-attention head (sah), cross-attention head (rah) and RTB one by one.}
\label{fig5}
\end{figure*}

We conduct ablation studies to better understand the impact of each component of our network. First, results of several encoding architectures are available to select the proper one as backbone. Then, considering the topology of vessels, regularization terms of loss function are discussed. Next, we analyze the effect of GTB based on the baseline, which is defined as the complete workflow without GTB and RTB. In order to identify whether vascular information contributes to lesion segmentation in advanced, we apply concatenation operation to the two outputs of GTBs directly. Finally, the roles of both self-attention head and cross-attention head in RTB are discussed thoroughly.
Since AUC\_PR is considered as the most important clinical evaluation metric for lesion segmentation in fundus images, we simplify the metrics to AUC\_PR only for ablative comparisons. The loss and PR curves obtained from different experiments are set out in Fig.\ref{fig5} and detailed AUC\_PR values can be compared in Table \ref{tab3}.

\subsubsection{Analysis on the Backbone}
To compare the effectiveness of backbone models, we perform experiments to select proper encoding architecture.    
ResNet-34\cite{he2016deep}, ResNet-50\cite{he2016deep}, Xception\cite{chollet2017xception}, Vgg19-bn\cite{simonyan2014very} and DenseNet-161\cite{huang2017densely} integrated with UNet are implemented with the segmentation heads of our network. As can be seen from the Table \ref{tab4}, the DenseNet-161 integrated with UNet achieves the best performance in all lesions and is utilized as the following backbone.

\begin{figure}[htbp]
\centerline{\includegraphics[width=\columnwidth]{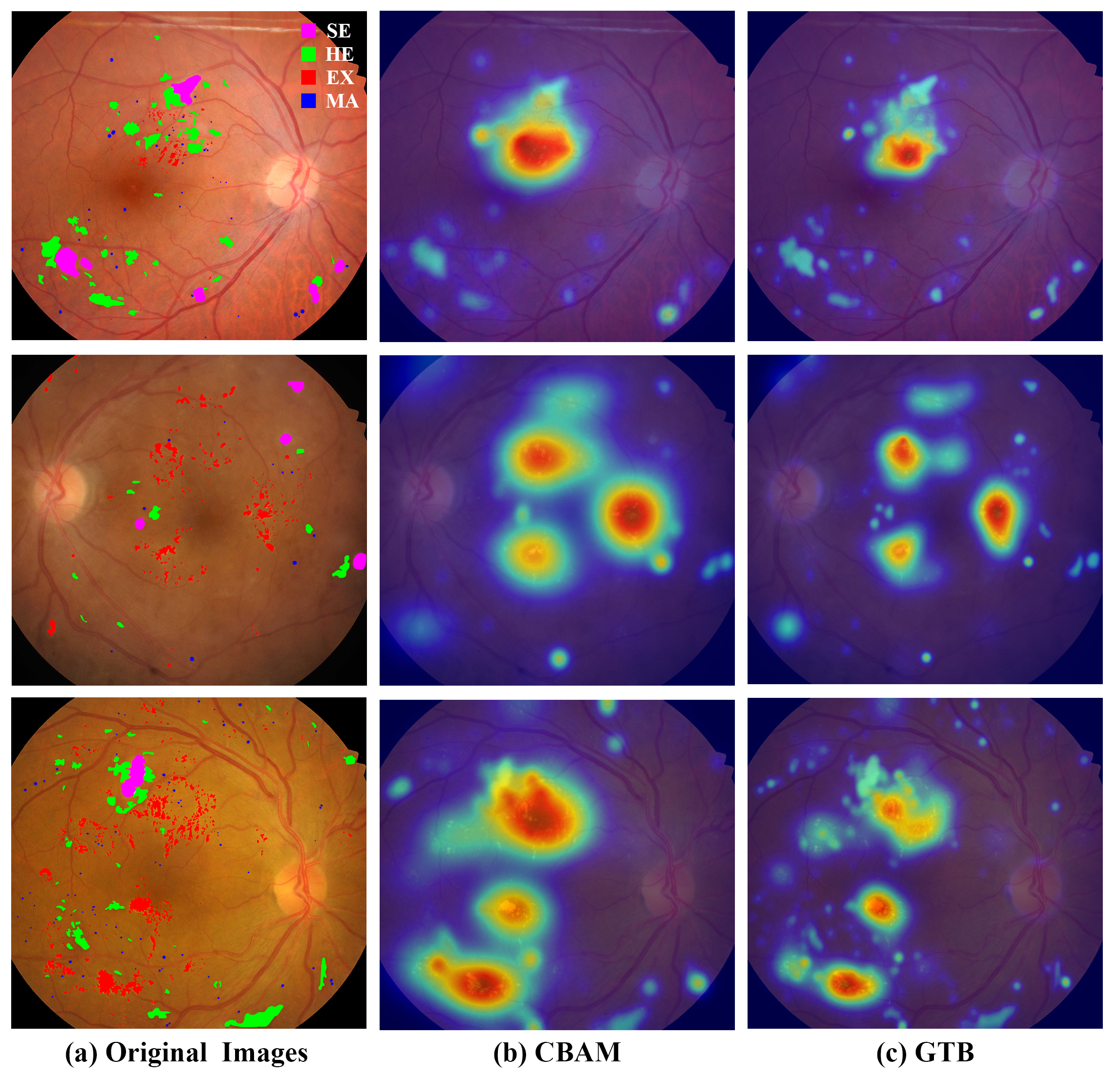}}
\caption{Visualization of (a) Original images with annotations; (b) spatial attention features with Convolutional Block Attention Module (CBAM) and (c) query attention features $\mathcal{F}$ with Global Transformer Block (GTB) for three images from IDRiD dataset. The GTB picks more discrete and small lesions up and fine-grains the patterns of interest.}
\label{fig6}
\end{figure}

\subsubsection{Analysis on the Regularization Term}
Considering the topology of vessels, an extension of loss function regularization terms has been conducted. In the result of vessel segmentation, there are two common cases, the neglected ends and the truncated trunks. 
Based on the above cases, we propose two regularization terms $R_{thin}$ and $R_{cl}$. The former inspired by \cite{yang2021hybrid} applies focal-loss function specifically on peripheral vessels, and the latter utilizes the center-line idea of \cite{shit2021cldice} to ensure the connectivity. Table \ref{tab_loss} indicates that the $R_{thin}$ improves the performance of MA segmentation and the $R_{cl}$ achieves the best grades in AUC\_PR of HE and MA. However, the alterations are unremarkable, probably due to the fact that the groundtruths of vessels are pseudo-masks generated by semi-supervision, rather than manually annotated masks. The upper bound on the performance of the vessel segmentation restricts the improvement of the regularization terms on the final results.

\begin{table}[]
  \centering
  \caption{Performance Comparison of Different Regularization Terms on IDRiD Dataset}
    \begin{tabular}{|c|c|c|c|c|c|}
    \hline
    \multirowcell{2}{$R_{thin}$}  &\multirowcell{2}{$R_{cl}$}  & \multicolumn{4}{c|}{AUC\_PR}\\\cline{3-6}
 &    & EX & HE & MA & SE \\
    \hline
&&\textbf{0.9024}    & 0.6880   & 0.4897   & \textbf{0.7502}\\
\checkmark&&0.8975&0.6845&0.4899&0.7432\\
&\checkmark&0.8912&\textbf{0.6891}&\textbf{0.4901}&0.7435\\
\checkmark&\checkmark&0.8923&0.6808&0.4895&0.7426\\
    \hline
    \end{tabular}
  \label{tab_loss}
\end{table}

\begin{figure*}[htbp]
\centerline{\includegraphics[width=\textwidth]{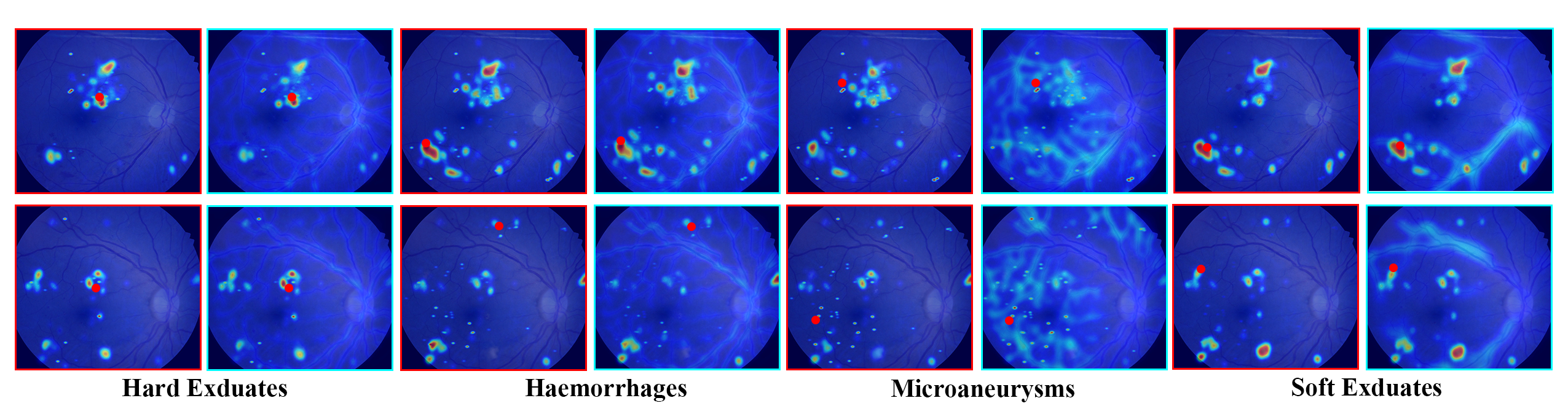}}
\caption{Visualization of the query position (red points) of different lesions and their two query-specific attention maps with Relation Transformer Block (RTB). The red borders denote the self-attention maps, and the blue denote the cross-attention maps. The attention of different query positions in EX, HE, MA and SE varies.}
\label{fig7}
\end{figure*}

\begin{table*}
    \centering
    \caption{Performance Comparison of different attention blocks on the IDRiD dataset}
    \label{tab5}
    \begin{tabular}{|c|c|c|c|c|c|c|c|c|}
    \hline
        lesion & \multicolumn{2}{|c|}{Hard   Exudates} &\multicolumn{2}{|c|}{Haemorrhages} 
        & \multicolumn{2}{|c|}{Microaneurysms} 
        & \multicolumn{2}{|c|}{Soft   Exudates}  \\ \hline
        method & AUC\_PR & AUC\_ROC & AUC\_PR & AUC\_ROC & AUC\_PR & AUC\_ROC & AUC\_PR & AUC\_ROC \\ \hline
        baseline(without CLAHE) & 0.8025 & 0.9912 & 0.6031 & 0.9498 & 0.3912 & 0.9803 & 0.5478 & 0.9120\\ \hline
        baseline & 0.8593  & 0.9919  & 0.6284  & 0.9553  & \textbf{0.4279}  & 0.9830  & 0.5766  & 0.9171  \\ 
        baseline+SENet\cite{hu2018squeeze} & 0.8653  & 0.9931  & 0.6408  & 0.9497  & 0.3861  & 0.9869  & 0.5683  & 0.9299  \\ 
        baseline+CBAM\cite{woo2018cbam} & 0.8606  & 0.9918  & 0.6470  & 0.9480  & 0.3979  & 0.9871  & 0.5525  & 0.9373  \\ 
        baseline+GC\cite{GCNet} & 0.8633  & 0.9929  & 0.6406  & 0.9488  & 0.4031  & 0.9809  & 0.5540  & 0.9251  \\ 
        baseline+GTB(Ours) & \textbf{0.8659}  & \textbf{0.9933}  & \textbf{0.6570}  & \textbf{0.9534}  & 0.4071  & \textbf{0.9879}  & \textbf{0.5968}  & \textbf{0.9458}  \\ \hline
    \end{tabular}
\end{table*}

\begin{table*}[]
  \centering
  \caption{Performance Comparison of different components of our network on the IDRiD dataset, where \textbf{cat} denotes a simple concatenate of multi-lesion and vessel features in channel-wise, \textbf{cah} and \textbf{sah} is abbreviations for cross-attention head and self-attention head respectively}
    \begin{tabular}{|c|c|c|c|c|c|c|c|c|}
    \hline
    \multirowcell{2}{Framework} & \multirowcell{2}{GTB} &\multirowcell{2}{cat} & \multicolumn{2}{c|}{RTB} & \multicolumn{4}{c|}{AUC\_PR}\\\cline{4-9}
     & & & cah & sah & EX & HE & MA & SE \\
    \hline
    \multirowcell{6}{baseline}     &       &       &       &       & 0.8593    & 0.6284  & 0.4071   & 0.5766   \\
          & \checkmark     &       &       &       & 0.8659   & 0.6570   & 0.4279   & 0.5968   \\
          & \checkmark     & \checkmark     &       &       & 0.8672    & 0.6756  & 0.4294    & 0.6663    \\
          & \checkmark     &       & \checkmark     &       & 0.8682    & 0.6818   & 0.4847    & 0.7463   \\
          & \checkmark     &       &       & \checkmark     & 0.8862    & 0.6846    & 0.4663   & 0.7422   \\
          & \checkmark     &       & \checkmark     & \checkmark     & \textbf{0.9024}  & \textbf{0.6880} 
        & \textbf{0.4897}  & \textbf{0.7502}  \\
    \hline
    \end{tabular}
  \label{tab3}
\end{table*}

\subsubsection{Analyze the Effect of GTB}

Table \ref{tab3} shows that GTB improves the performance on the basis of baseline, which indicates the weights assigned to channels in each position by GTB benefit the segmentation of all four lesions. To highlight the performance of GTB further, we compare it with other popular attention blocks under the same model parameters. 
Table \ref{tab5} illustrates that the performance of GTB on IDRiD dataset is better than others. Different from the pooling operation as other attention blocks do, channel-wise weights of GTB are specific in each pixel, so that different channels are enlightened in different pixels.

As shown in Fig.\ref{fig6}, the snapshots of attentive features after softmax function are taken in color. In the comparison of the spatial attention features of CBAM\cite{woo2018cbam} and the attention features $\mathcal{F}$ of GTB, many discrete and small patterns overlooked by the former are noticed by the latter.

\subsubsection{Analyze the Effect of RTB}
Table \ref{tab3} lists the ablation results. We first reaffirm the idea that vascular information contributes to multi-lesion segmentation by concatenating the vascular and multi-lesion features in channel-wise. In the comparison with and without concatenations, the greatest performance gains are obtained for HE and SE, which is consistent with the previous analysis that the HE and SE have strong relations with vessels. In order to make full use of vascular information, RTB is applied instead of simple concatenation.
Compared with the simple concatenation, cross-attention head incorporating multi-lesion and vascular features in a transformer way enhances the scores of all four lesions further. Likewise, with the integration of self-attention head, the scores get a huge improvement as well. 

As visualized in Fig.\ref{fig7}, query-specific attention maps focus on specialized tissues. Take the query pixel on MA as an example, the self-attention tends to smaller patterns, and the cross-attention specializes the vascular tributaries, which are supposed to assist in reducing false negatives mistaken for tributaries and eliminating false alarms far from tributaries.
Both self-attention head and cross-attention head play a role in improving the network performance, evincing the fact that exploring the internal relationships of multi-lesion and vessels makes sense.

Finally both GTB and RTB are incorporated and the network achieves the highest results. As shown in the bottom half of the Fig.\ref{fig5}, the curve corresponding to the complete network wraps almost entirely around the others.

\begin{figure*}[htbp]
\centerline{\includegraphics[width=\textwidth]{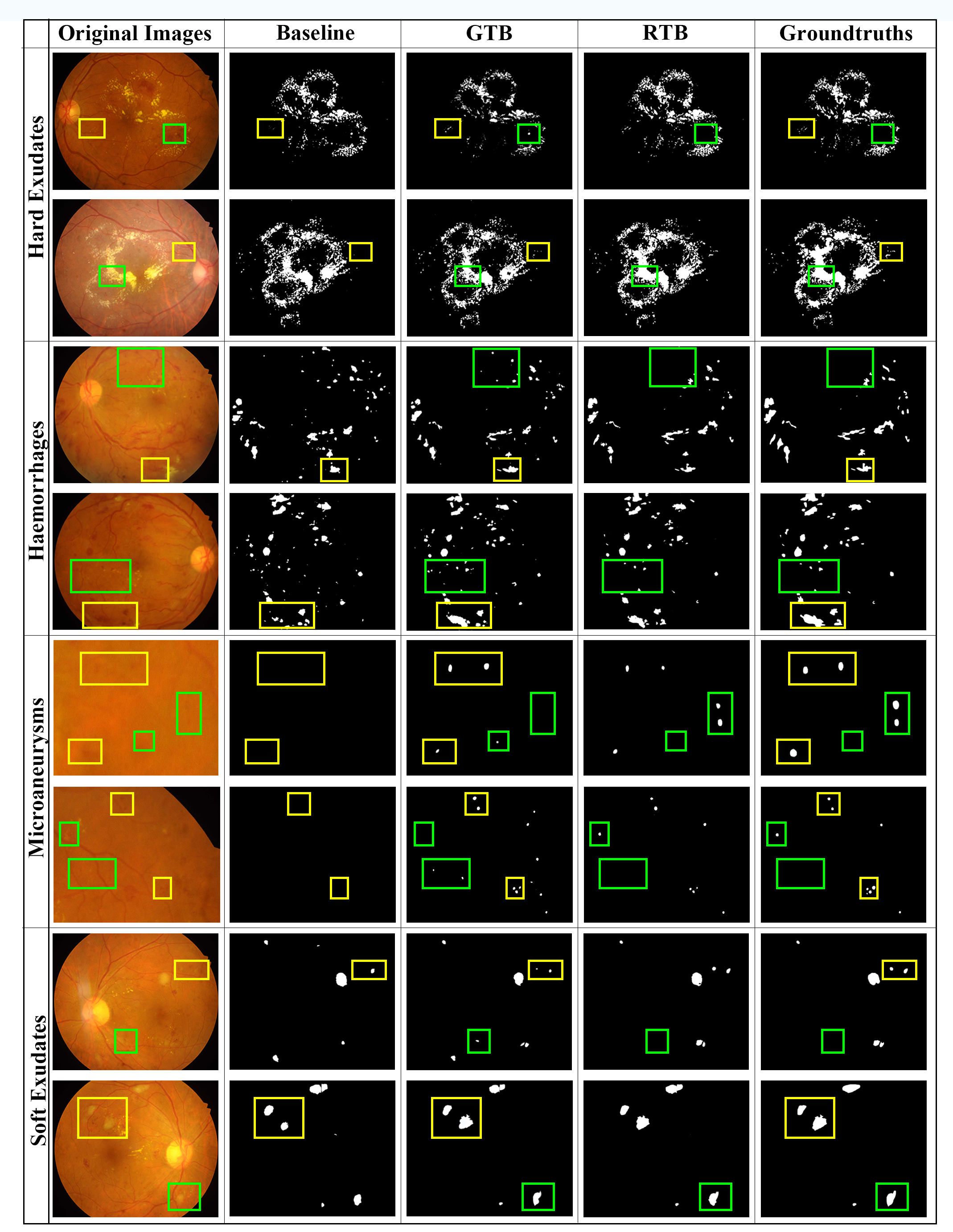}}
\caption{Visualization of segmentation results for multi-lesion segmentation on IDRiD dataset. The different columns represent the original images, the segmentation results generated by baseline, baseline+GTB, baseline+GTB+RTB and groundtruths respectively. The yellow boxes denote the improvements over baseline brought about by GTB, in the form of pickups of missing detections, while the green boxes denote the improvements over baseline+GTB brought about by RTB, mainly in the form of fewer false alarms.
}
\label{fig8}
\end{figure*}

\subsection{Generalization Studies on DDR and IDRiD Dataset}
For medical images, it is challenging but meaningful to realize the generalization over different domains under different imaging conditions. In purpose to validate the generalization capability, models are trained with the images from the train set of DDR dataset and tested on test set of IDRiD dataset which is captured from another source. Table \ref{tab6} compares the results obtained from the preliminary analysis of generalization. From the chart, it can be seen that our method achieves the best performance by narrowing down the gap between images under different conditions.

\subsection{Qualitative Results}
To better illustrate the effect of GTB and RTB, we visualize the results of certain images. 
Fig.\ref{fig8} compares the segmentation results with corresponding original images and groundtruths. We take the segmentation maps of baseline, baseline with GTB and baseline with GTB and RTB to present the improvements of different components of our network.
The yellow boxes present the improvements of GTB, where the missing detections are discovered. The green boxes are steps up from the yellow boxes by RTB, which picks up missing detections further and reduces false alarms. Additionally, the edges of the large lesion patterns are more precisely fine-tuned by RTB, especially for the SE with blurred edges.

\begin{table}[]
  \centering
  \caption{Performance Comparison of different methods on the generalization from DDR dataset to IDRiD dataset}
    \begin{tabular}{|c|c|c|c|c|}
    \hline
    \multirowcell{2}{Framework}  & \multicolumn{4}{c|}{AUC\_PR}\\\cline{2-5}
     & EX & HE & MA & SE \\
    \hline
     HED\cite{xie2015holistically} & 0.5420    & 0.2104  & 0.1245   & 0.1278   \\
     DeepLab v3+\cite{chen2018deeplabv3} & 0.6480    & 0.4472   & 0.1823   & 0.2926   \\
     UNet\cite{guan2019fully,Yakubovskiy2019} & 0.6472 & 0.4452 & 0.1965 & 0.2845 \\
     Lseg\cite{guo2019seg} & 0.6501    & 0.4405  & 0.1986    & 0.3059    \\
     RTN(Ours) & \textbf{0.6799}    & \textbf{0.4504}   & \textbf{0.2114}    & \textbf{0.3401}   \\
    \hline
    \end{tabular}
  \label{tab6}
\end{table}

\section{Conclusion And Discussion}
In this paper, we present a novel network that employs a dual-branch architecture with GTB and RTB to segment the four DR lesions simultaneously. 
Outstanding experiment results of our network can be attributed to GTB and RTB, which investigate the intra-class dependencies among multi-lesion and inter-class relations of multi-lesion and vessels.

However, limited to the considerable cost of expertise pixel-level annotations, the vessel pseudo masks provided by semi-supervised learning are inevitably coarse-grained and lead to the inadequacy of our network. Therefore,
in our future work, we will further modify the vascular semi-supervised learning  strategy and keep improving the transformer structures to achieve better performance in DR multi-lesion segmentation with less memory requirement.


\bibliographystyle{IEEEtran}
\bibliography{ref}

\begin{thebibliography}{10}
\providecommand{\url}[1]{#1}
\csname url@samestyle\endcsname
\providecommand{\newblock}{\relax}
\providecommand{\bibinfo}[2]{#2}
\providecommand{\BIBentrySTDinterwordspacing}{\spaceskip=0pt\relax}
\providecommand{\BIBentryALTinterwordstretchfactor}{4}
\providecommand{\BIBentryALTinterwordspacing}{\spaceskip=\fontdimen2\font plus
\BIBentryALTinterwordstretchfactor\fontdimen3\font minus
  \fontdimen4\font\relax}
\providecommand{\BIBforeignlanguage}[2]{{%
\expandafter\ifx\csname l@#1\endcsname\relax
\typeout{** WARNING: IEEEtran.bst: No hyphenation pattern has been}%
\typeout{** loaded for the language `#1'. Using the pattern for}%
\typeout{** the default language instead.}%
\else
\language=\csname l@#1\endcsname
\fi
#2}}
\providecommand{\BIBdecl}{\relax}
\BIBdecl

\bibitem{thomas2019idf}
R.~Thomas, S.~Halim, S.~Gurudas, S.~Sivaprasad, and D.~Owens, ``Idf diabetes
  atlas: A review of studies utilising retinal photography on the global
  prevalence of diabetes related retinopathy between 2015 and 2018,''
  \emph{Diabetes research and clinical practice}, vol. 157, p. 107840, 2019.

\bibitem{ciulla2003diabetic}
T.~A. Ciulla, A.~G. Amador, and B.~Zinman, ``Diabetic retinopathy and diabetic
  macular edema: pathophysiology, screening, and novel therapies,''
  \emph{Diabetes care}, vol.~26, no.~9, pp. 2653--2664, 2003.

\bibitem{raman2016diabetic}
R.~Raman, L.~Gella, S.~Srinivasan, and T.~Sharma, ``Diabetic retinopathy: An
  epidemic at home and around the world,'' \emph{Indian journal of
  ophthalmology}, vol.~64, no.~1, p.~69, 2016.

\bibitem{wong2018guidelines}
T.~Y. Wong, J.~Sun, R.~Kawasaki, P.~Ruamviboonsuk, N.~Gupta, V.~C. Lansingh,
  M.~Maia, W.~Mathenge, S.~Moreker, M.~M. Muqit \emph{et~al.}, ``Guidelines on
  diabetic eye care: the international council of ophthalmology recommendations
  for screening, follow-up, referral, and treatment based on resource
  settings,'' \emph{Ophthalmology}, vol. 125, no.~10, pp. 1608--1622, 2018.

\bibitem{ting2016diabetic}
D.~S.~W. Ting, G.~C.~M. Cheung, and T.~Y. Wong, ``Diabetic retinopathy: global
  prevalence, major risk factors, screening practices and public health
  challenges: a review,'' \emph{Clinical \& experimental ophthalmology},
  vol.~44, no.~4, pp. 260--277, 2016.

\bibitem{guo2019seg}
S.~Guo, T.~Li, H.~Kang, N.~Li, Y.~Zhang, and K.~Wang, ``L-seg: An end-to-end
  unified framework for multi-lesion segmentation of fundus images,''
  \emph{Neurocomputing}, vol. 349, pp. 52--63, 2019.

\bibitem{simonyan2014very}
K.~Simonyan and A.~Zisserman, ``Very deep convolutional networks for
  large-scale image recognition,'' \emph{arXiv:1409.1556}, 2014.

\bibitem{zhou2019collaborative}
Y.~Zhou, X.~He, L.~Huang, L.~Liu, F.~Zhu, S.~Cui, and L.~Shao, ``Collaborative
  learning of semi-supervised segmentation and classification for medical
  images,'' in \emph{Proceedings of the IEEE/CVF Conference on Computer Vision
  and Pattern Recognition (CVPR)}, 2019, pp. 2079--2088.

\bibitem{tavakoli2020automated}
M.~Tavakoli, S.~Jazani, and M.~Nazar, ``Automated detection of microaneurysms
  in color fundus images using deep learning with different preprocessing
  approaches,'' in \emph{Medical Imaging 2020: Imaging Informatics for
  Healthcare, Research, and Applications}, vol. 11318.\hskip 1em plus 0.5em
  minus 0.4em\relax International Society for Optics and Photonics, 2020, p.
  113180E.

\bibitem{mamilla2017extraction}
R.~T. Mamilla, V.~K.~R. Ede, and P.~R. Bhima, ``Extraction of microaneurysms
  and hemorrhages from digital retinal images,'' \emph{Journal of Medical and
  Biological Engineering}, vol.~37, no.~3, pp. 395--408, 2017.

\bibitem{wu2017automatic}
B.~Wu, W.~Zhu, F.~Shi, S.~Zhu, and X.~Chen, ``Automatic detection of
  microaneurysms in retinal fundus images,'' \emph{Computerized Medical Imaging
  and Graphics}, vol.~55, pp. 106--112, 2017.

\bibitem{khojasteh2019novel}
P.~Khojasteh, B.~Aliahmad, and D.~K. Kumar, ``A novel color space of fundus
  images for automatic exudates detection,'' \emph{Biomedical Signal Processing
  and Control}, vol.~49, pp. 240--249, 2019.

\bibitem{mo2018exudate}
J.~Mo, L.~Zhang, and Y.~Feng, ``Exudate-based diabetic macular edema
  recognition in retinal images using cascaded deep residual networks,''
  \emph{Neurocomputing}, vol. 290, pp. 161--171, 2018.

\bibitem{GCNet}
J.~Ni, J.~Wu, J.~Tong, Z.~Chen, and J.~Zhao, ``Gc-net: Global context network
  for medical image segmentation,'' \emph{Computer Methods and Programs in
  Biomedicine}, vol. 190, p. 105121, 2019.

\bibitem{walter2007automatic}
T.~Walter, P.~Massin, A.~Erginay, R.~Ordonez, C.~Jeulin, and J.-C. Klein,
  ``Automatic detection of microaneurysms in color fundus images,''
  \emph{Medical Image Analysis}, vol.~11, no.~6, pp. 555--566, 2007.

\bibitem{Automatic2005}
M.~Niemeijer, B.~van Ginneken, J.~Staal, M.~Suttorp-Schulten, and M.~Abramoff,
  ``{Automatic detection of red lesions in digital color fundus photographs},''
  \emph{{IEEE Transactions on Medical Imaging}}, vol.~{24}, no.~{5}, pp.
  {584--592}, {2005}.

\bibitem{alipour2012analysis}
S.~H.~M. Alipour, H.~Rabbani, M.~Akhlaghi, A.~M. Dehnavi, and S.~H. Javanmard,
  ``Analysis of foveal avascular zone for grading of diabetic retinopathy
  severity based on curvelet transform,'' \emph{Graefe's Archive for Clinical
  and Experimental Ophthalmology}, vol. 250, no.~11, pp. 1607--1614, 2012.

\bibitem{krizhevsky2012imagenet}
A.~Krizhevsky, I.~Sutskever, and G.~E. Hinton, ``Imagenet classification with
  deep convolutional neural networks,'' \emph{Advances in Neural Information
  Processing Systems}, vol.~25, pp. 1097--1105, 2012.

\bibitem{CABNet}
A.~He, T.~Li, N.~Li, K.~Wang, and H.~Fu, ``{CABNet: Category Attention Block
  for Imbalanced Diabetic Retinopathy Grading},'' \emph{{IEEE Transactions on
  Medical Imaging}}, vol.~{40}, no.~{1}, pp. {143--153}, {2021}.

\bibitem{CANet}
X.~Li, X.~Hu, L.~Yu, L.~Zhu, C.-W. Fu, and P.-A. Heng, ``{CANet: Cross-Disease
  Attention Network for Joint Diabetic Retinopathy and Diabetic Macular Edema
  Grading},'' \emph{{IEEE Transactions on Medical Imaging}}, vol.~{39},
  no.~{5}, pp. {1483--1493}, {2020}.

\bibitem{CENet}
Z.~Gu, J.~Cheng, H.~Fu, K.~Zhou, H.~Hao, Y.~Zhao, T.~Zhang, S.~Gao, and J.~Liu,
  ``{CE-Net: Context Encoder Network for 2D Medical Image Segmentation},''
  \emph{{IEEE Transactions on Medical Imaging}}, vol.~{38}, no.~{10}, pp.
  {2281--2292}, {2019}.

\bibitem{zhang2019detection}
L.~Zhang, S.~Feng, G.~Duan, Y.~Li, and G.~Liu, ``Detection of microaneurysms in
  fundus images based on an attention mechanism,'' \emph{Genes}, vol.~10,
  no.~10, p. 817, 2019.

\bibitem{wang2017zoom}
Z.~Wang, Y.~Yin, J.~Shi, W.~Fang, H.~Li, and X.~Wang, ``Zoom-in-net: Deep
  mining lesions for diabetic retinopathy detection,'' in \emph{International
  Conference on Medical Image Computing and Computer-Assisted Intervention
  (MICCAI)}.\hskip 1em plus 0.5em minus 0.4em\relax Springer, 2017, pp.
  267--275.

\bibitem{xie2020sesv}
Y.~Xie, J.~Zhang, H.~Lu, C.~Shen, and Y.~Xia, ``Sesv: Accurate medical image
  segmentation by predicting and correcting errors,'' \emph{IEEE Transactions
  on Medical Imaging}, vol.~40, no.~1, pp. 286--296, 2020.

\bibitem{vaswani2017attention}
A.~Vaswani, N.~Shazeer, N.~Parmar, J.~Uszkoreit, L.~Jones, A.~N. Gomez,
  L.~Kaiser, and I.~Polosukhin, ``Attention is all you need,''
  \emph{arXiv:1706.03762}, 2017.

\bibitem{dosovitskiy2020image}
A.~Dosovitskiy, L.~Beyer, A.~Kolesnikov, D.~Weissenborn, X.~Zhai,
  T.~Unterthiner, M.~Dehghani, M.~Minderer, G.~Heigold, S.~Gelly \emph{et~al.},
  ``An image is worth 16x16 words: Transformers for image recognition at
  scale,'' \emph{arXiv:2010.11929}, 2020.

\bibitem{GPT}
Y.~Liu, H.~Yuan, Z.~Wang, and S.~Ji, ``Global pixel transformers for virtual
  staining of microscopy images,'' \emph{IEEE Transactions on Medical Imaging},
  vol.~39, no.~6, pp. 2256--2266, 2020.

\bibitem{guo2021transformer}
D.~Guo and D.~Terzopoulos, ``A transformer-based network for anisotropic 3d
  medical image segmentation,'' in \emph{Proceedings of International
  Conference on Patern Recognition (ICPR)}.\hskip 1em plus 0.5em minus
  0.4em\relax IEEE, 2021, pp. 8857--8861.

\bibitem{DRT}
D.~Song, B.~Fu, F.~Li, J.~Xiong, J.~He, X.~Zhang, and Y.~Qiao, ``Deep relation
  transformer for diagnosing glaucoma with optical coherence tomography and
  visual field function,'' \emph{IEEE Transactions on Medical Imaging}, 2021.

\bibitem{wang2018NLnet}
X.~Wang, R.~Girshick, A.~Gupta, and K.~He, ``Non-local neural networks,'' in
  \emph{Proceedings of the IEEE Conference on Computer Vision and Pattern
  Recognition (CVPR)}, 2018, pp. 7794--7803.

\bibitem{porwal2018indian}
P.~Porwal, S.~Pachade, R.~Kamble, M.~Kokare, G.~Deshmukh, V.~Sahasrabuddhe, and
  F.~Meriaudeau, ``Indian diabetic retinopathy image dataset (idrid): a
  database for diabetic retinopathy screening research,'' \emph{Data}, vol.~3,
  no.~3, p.~25, 2018.

\bibitem{porwal2020idrid}
P.~Porwal, S.~Pachade, M.~Kokare, G.~Deshmukh, J.~Son, W.~Bae, L.~Liu, J.~Wang,
  X.~Liu, L.~Gao \emph{et~al.}, ``Idrid: Diabetic retinopathy--segmentation and
  grading challenge,'' \emph{Medical Image Analysis}, vol.~59, p. 101561, 2020.

\bibitem{LI2019}
\BIBentryALTinterwordspacing
T.~Li, Y.~Gao, K.~Wang, S.~Guo, H.~Liu, and H.~Kang, ``Diagnostic assessment of
  deep learning algorithms for diabetic retinopathy screening,''
  \emph{Information Sciences}, vol. 501, pp. 511 -- 522, 2019. [Online].
  Available:
  \url{http://www.sciencedirect.com/science/article/pii/S0020025519305377}
\BIBentrySTDinterwordspacing

\bibitem{kou2019microaneurysms}
C.~Kou, W.~Li, W.~Liang, Z.~Yu, and J.~Hao, ``Microaneurysms segmentation with
  a u-net based on recurrent residual convolutional neural network,''
  \emph{Journal of Medical Imaging}, vol.~6, no.~2, p. 025008, 2019.

\bibitem{xie2015holistically}
S.~Xie and Z.~Tu, ``Holistically-nested edge detection,'' in \emph{Proceedings
  of the IEEE International Conference on Computer Vision (ICCV)}, 2015, pp.
  1395--1403.

\bibitem{chen2018deeplabv3}
L.-C. Chen, Y.~Zhu, G.~Papandreou, F.~Schroff, and H.~Adam, ``Encoder-decoder
  with atrous separable convolution for semantic image segmentation,'' in
  \emph{Proceedings of the European Conference on Computer Vision (ECCV)},
  2018, pp. 801--818.

\bibitem{guan2019fully}
S.~Guan, A.~A. Khan, S.~Sikdar, and P.~V. Chitnis, ``Fully dense unet for 2-d
  sparse photoacoustic tomography artifact removal,'' \emph{IEEE journal of
  biomedical and health informatics}, vol.~24, no.~2, pp. 568--576, 2019.

\bibitem{Yakubovskiy2019}
\BIBentryALTinterwordspacing
P.~Yakubovskiy, ``Segmentation models pytorch,'' 2020. [Online]. Available:
  \url{https://github.com/qubvel/segmentation\_models.pytorch}
\BIBentrySTDinterwordspacing

\bibitem{staal2004ridgeb91}
J.~Staal, M.~D. Abr{\`a}moff, M.~Niemeijer, M.~A. Viergever, and
  B.~Van~Ginneken, ``Ridge-based vessel segmentation in color images of the
  retina,'' \emph{IEEE Transactions on Medical Imaging}, vol.~23, no.~4, pp.
  501--509, 2004.

\bibitem{hoover2000locatingb92}
A.~Hoover, V.~Kouznetsova, and M.~Goldbaum, ``Locating blood vessels in retinal
  images by piecewise threshold probing of a matched filter response,''
  \emph{IEEE Transactions on Medical Imaging}, vol.~19, no.~3, pp. 203--210,
  2000.

\bibitem{huang2017densely}
G.~Huang, Z.~Liu, L.~Van Der~Maaten, and K.~Q. Weinberger, ``Densely connected
  convolutional networks,'' in \emph{Proceedings of the IEEE conference on
  computer vision and pattern recognition (CVPR)}, 2017, pp. 4700--4708.

\bibitem{scikit-learn}
F.~Pedregosa, G.~Varoquaux, A.~Gramfort, V.~Michel, B.~Thirion, O.~Grisel,
  M.~Blondel, P.~Prettenhofer, R.~Weiss, V.~Dubourg, J.~Vanderplas, A.~Passos,
  D.~Cournapeau, M.~Brucher, M.~Perrot, and E.~Duchesnay, ``Scikit-learn:
  Machine learning in {P}ython,'' \emph{Journal of Machine Learning Research},
  vol.~12, pp. 2825--2830, 2011.

\bibitem{he2016deep}
K.~He, X.~Zhang, S.~Ren, and J.~Sun, ``Deep residual learning for image
  recognition,'' in \emph{Proceedings of the IEEE conference on computer vision
  and pattern recognition (CVPR)}, 2016, pp. 770--778.

\bibitem{chollet2017xception}
F.~Chollet, ``Xception: Deep learning with depthwise separable convolutions,''
  in \emph{Proceedings of the IEEE conference on computer vision and pattern
  recognition (CVPR)}, 2017, pp. 1251--1258.

\bibitem{yang2021hybrid}
L.~Yang, H.~Wang, Q.~Zeng, Y.~Liu, and G.~Bian, ``A hybrid deep segmentation
  network for fundus vessels via deep-learning framework,''
  \emph{Neurocomputing}, vol. 448, pp. 168--178, 2021.

\bibitem{shit2021cldice}
S.~Shit, J.~C. Paetzold, A.~Sekuboyina, I.~Ezhov, A.~Unger, A.~Zhylka, J.~P.
  Pluim, U.~Bauer, and B.~H. Menze, ``cldice-a novel topology-preserving loss
  function for tubular structure segmentation,'' in \emph{Proceedings of the
  IEEE international conference on computer vision (CVPR)}, 2021, pp.
  16\,560--16\,569.

\bibitem{hu2018squeeze}
J.~Hu, L.~Shen, and G.~Sun, ``Squeeze-and-excitation networks,'' in
  \emph{Proceedings of the IEEE Conference on Computer Vision and Pattern
  Recognition (CVPR)}, 2018, pp. 7132--7141.

\bibitem{woo2018cbam}
S.~Woo, J.~Park, J.-Y. Lee, and I.~S. Kweon, ``Cbam: Convolutional block
  attention module,'' in \emph{Proceedings of the European Conference on
  Computer Vision (ECCV)}, 2018, pp. 3--19.

\end{thebibliography}
\end{document}